\documentclass{moriond}
\usepackage{amsmath}
\usepackage[dvipsnames]{xcolor}
\usepackage{xspace}
\usepackage{upgreek}
\usepackage{lineno}
\usepackage{scalerel}

\def\Journal#1#2#3#4{{#1} {\bf #2}, #3 (#4)}

\def\PRL{\em Phys. Rev. Lett.}
\def\PRD{{\em Phys. Rev.} D}

\def\be{\begin{equation}}
\def\ee{\end{equation}}
\def\bea{\begin{eqnarray}}
\def\eea{\end{eqnarray}}

\def\thebaroffset{0.18em}
\newcommand{\offsetoverline}[2][\thebaroffset]{\kern #1\overline{\kern -#1 #2}}%
\def\PB      {\ensuremath{B}\xspace}   
\def\B       {{\ensuremath{\PB}}\xspace}
\def\Bbar    {{\ensuremath{\offsetoverline{\PB}}}\xspace}
\def\PK      {\ensuremath{K}\xspace}
\def\Pmu         {\ensuremath{\mu}\xspace}
\def\kaon    {{\ensuremath{\PK}}\xspace}

\def\ellell     {\ensuremath{\ell \ell}\xspace}
\def\Kstarz  {{\ensuremath{\kaon^{*0}}}\xspace}

\def\mup        {{\ensuremath{\Pmu^+}}\xspace}
\def\mun        {{\ensuremath{\Pmu^-}}\xspace}

\def\Bd      {{\ensuremath{\B^0}}\xspace}
\def\Bdb     {{\ensuremath{\Bbar{}^0}}\xspace}
\def\qsq{{\ensuremath{q^2}}\xspace}
\def\mkpi{\ensuremath{m_{\kaon\pi}}\xspace}
\def\mkpimumu{\ensuremath{m_{\kaon\pi\Pmu\Pmu}}\xspace}
\def\lhcb{\mbox{LHCb}\xspace}
\newcommand{\decay}[2]{\ensuremath{#1\!\to #2}\xspace} 
\def\BdToKstmm{\decay{\Bd}{\Kstarz\mup\mun}}
\def\Pb      {\ensuremath{b}\xspace} 
\def\Ps      {\ensuremath{s}\xspace} 
\def\Pc      {\ensuremath{c}\xspace} 
\def\squark    {{\ensuremath{\Ps}}\xspace}
\def\cquark    {{\ensuremath{\Pc}}\xspace}
\def\cbarquark {{\ensuremath{\overline \cquark}}\xspace}
\def\bquark    {{\ensuremath{\Pb}}\xspace}
\def\bTosll{\decay{\bquark}{\squark\ellell}}

\newcommand{\aunit}[1]{\ensuremath{\text{\,#1}}}  
\def\fb   {\ensuremath{\aunit{fb}}\xspace}
\def\invfb   {\ensuremath{\fb^{-1}}\xspace}

\newcommand{\gev}{\aunit{Ge\kern -0.1em V}\xspace}
\newcommand{\mev}{\aunit{Me\kern -0.1em V}\xspace}
 
\newcommand{\mevc}{\ensuremath{\aunit{Me\kern -0.1em V\!/}c}\xspace}
\newcommand{\gevc}{\ensuremath{\aunit{Ge\kern -0.1em V\!/}c}\xspace}
\newcommand{\mevcc}{\ensuremath{\aunit{Me\kern -0.1em V\!/}c^2}\xspace}
\newcommand{\gevcc}{\ensuremath{\aunit{Ge\kern -0.1em V\!/}c^2}\xspace}
\newcommand{\gevgevcccc}{\ensuremath{\gev^2\!/c^4}\xspace} 
\def\CP                {{\ensuremath{C\!P}}\xspace}
\newcommand{\ACP}{{\ensuremath{{\mathcal{A}}^{\CP}}}\xspace}

\def\thetal       {\ensuremath{\theta_\ell}\xspace}
\def\thetak       {\ensuremath{\theta_K}\xspace}
\def\phiangular       {\ensuremath{\phi}\xspace}

\newcommand{\Photo}{}

\begin{document}

\vspace*{4cm}
\title{Binned angular analysis of \BdToKstmm with \lhcb}

\author{ Leon Carus, on behalf of the \lhcb collaboration }

\address{Physikalisches Institut, Universität Heidelberg, Im Neuenheimer Feld 226,\\ 69120 Heidelberg, Germany}

\maketitle\abstracts{
    The analysis of \bTosll flavor-changing neutral current decays is a powerful test of the Standard Model (SM).
    Due to the strong suppression of these modes in the SM, potential New Physics contributions can have a significant impact on the 
    measured physics observables.
    The angular analysis of \BdToKstmm decays gives access to optimized angular observables, which are less dependent on hadronic form factors than for example branching fraction measurements.
    The angular observables are extracted in bins of the invariant dimuon mass squared, \qsq, making the analysis model-independent with respect to any assumptions about the shape of the signal distribution in \qsq.
    Previous binned measurements of the \BdToKstmm angular observables by \lhcb were in tension with the SM at the level of $3.3$ standard deviations ($\sigma$).
    These proceedings present the ongoing effort to update the binned angular analysis to include the full Run~1 and Run~2 data sample of \lhcb. }

\section{Introduction}
\begin{figure}[h]
\begin{minipage}{0.49\linewidth}
 \centerline{\includegraphics[width=0.75\textwidth, height=3.3cm]{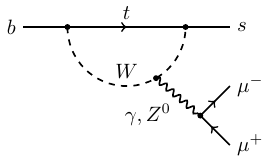}}
\end{minipage}
\hfill
\begin{minipage}{0.49\linewidth}
\centerline{\includegraphics[width=0.75\textwidth, height=3.3cm]{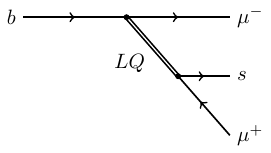}}
\end{minipage}
\vspace*{-2mm}
\caption{\label{Fig:feynman} Feynman diagram of the SM $b\rightarrow s \mu \mu$ penguin contribution (left).
    Example of NP tree-level decay including a lepto-quark. }

\end{figure}

\begin{figure}
\begin{minipage}{0.49\linewidth}
 \centerline{\includegraphics[width=1.0\textwidth]{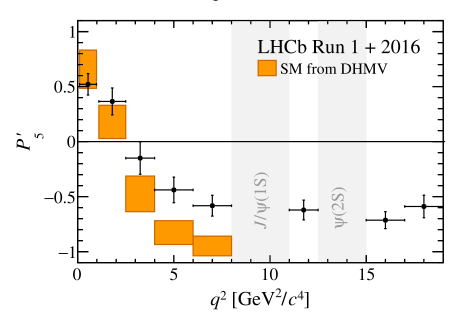}}
\end{minipage}
\hfill
\begin{minipage}{0.49\linewidth}
\vspace{-0.5cm}
\centerline{\includegraphics[width=1.0\textwidth, height=5.3cm]{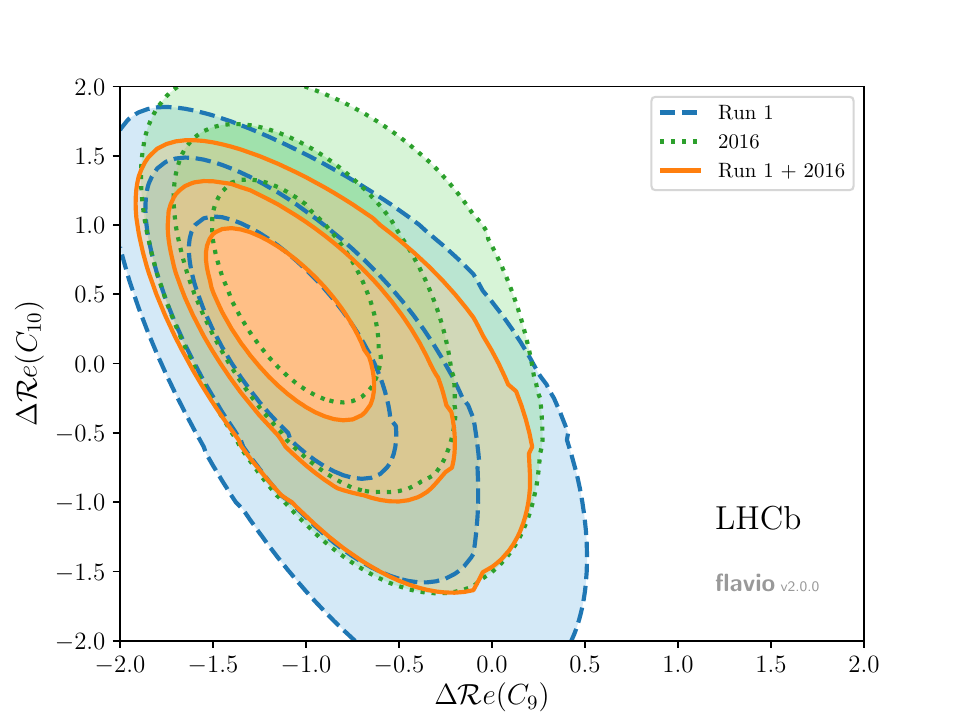}}
\end{minipage}
\vspace*{-2mm}
\caption{\label{Fig:prev_measurement} Measurement of the less form-factor dependent observable $P_5^\prime$ (left) by \lhcb using Run~1 and 2016 data in black.
    The SM predictions~\protect\cite{Descotes-Genon,Khodjamirian:2010vf} are displayed in orange. The right plot shows the results of the Wilson Coefficient shifts $\Delta {\ensuremath{\mathcal{R}e(\mathcal{C}_{9})}\xspace}$ and $\Delta {\ensuremath{\mathcal{R}e(\mathcal{C}_{10})}\xspace}$ derived from the measured angular observables. Both plots show the measurement to be in tension with the SM prediction.}

\end{figure}

The decay \BdToKstmm occurs via a \bTosll flavor-changing neutral current (FCNC) transition.
These transitions are not allowed in the SM at tree level and can therefore only occur as electroweak penguin decays at the loop level as displayed in Fig.~\ref{Fig:feynman} (left).
The combination of loop- and CKM-suppression leads to a very high sensitivity to New Physics (NP) effects, which can potentially contribute at the tree level as shown in Fig.~\ref{Fig:feynman} (right). 
Previous analysis of \bTosll decays by \lhcb found tensions of data with SM predictions in angular observables~\cite{LHCb-PAPER-2020-002,kstplus_angular,phi_angular} and branching fraction measurements~\cite{isospin,swave-paper,bsphimumu,lambda}. 
The angular analysis of the decay \BdToKstmm allows access to optimized observables, which are less dependent on the theoretical uncertainties of hadronic form factors~\cite{optimized_observables}.
The binned angular analysis of \BdToKstmm decays using the Run~1 and 2016 data sample of \lhcb, saw a global tension between the SM prediction and its measurement of about $3.3~\sigma$~\cite{LHCb-PAPER-2020-002}.
The results and predictions for $P_5^{\prime}$ and the Wilson Coefficient shifts $\Delta {\ensuremath{\mathcal{R}e(\mathcal{C}_{9})}\xspace}$ and $\Delta {\ensuremath{\mathcal{R}e(\mathcal{C}_{10})}\xspace}$ are shown in Fig.~\ref{Fig:prev_measurement}.
A consistent shift of $P_5^{\prime}$ to higher values is visible in the left plot. The Wilson Coefficients measurement prefers a shift from the SM value to ${\Delta \ensuremath{\mathcal{R}e(\mathcal{C}_{9})}\xspace}\approx -1$.

\newpage
The estimation of hadronic form factor uncertainties and the effect of \cquark\cbarquark-loop contributions present a significant challenge for theory predictions of the angular observables and are an active field of study.
In contrast to the \qsq-unbinned angular analysis~\cite{amplitudeanalysis,amplitudeanalysis2,dispersion_relation}, the measurement of binned angular observables is largely model-independent.
These proceedings discuss the currently ongoing effort to update the binned angular analysis of \BdToKstmm adding the \lhcb data samples from 2017 and 2018.
This increases the integrated luminosity to a total of 8.4 \invfb.

\section{Five-dimensional angular fit to \BdToKstmm}

\begin{figure}
\begin{minipage}{0.49\linewidth}
 \centerline{\includegraphics[width=1.0\textwidth, height=5.3cm]{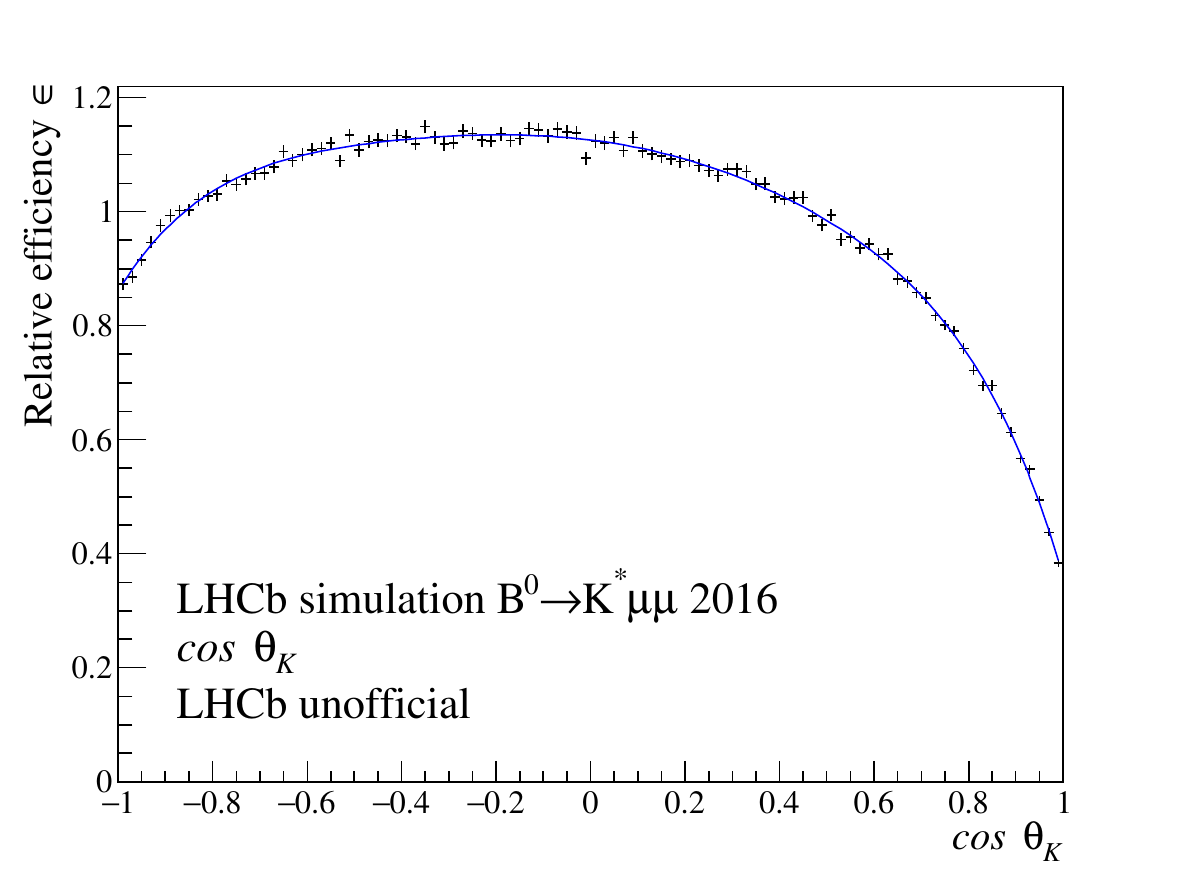}}
\end{minipage}
\hfill
\begin{minipage}{0.49\linewidth}
\centerline{\includegraphics[width=1.0\textwidth, height=5.3cm]{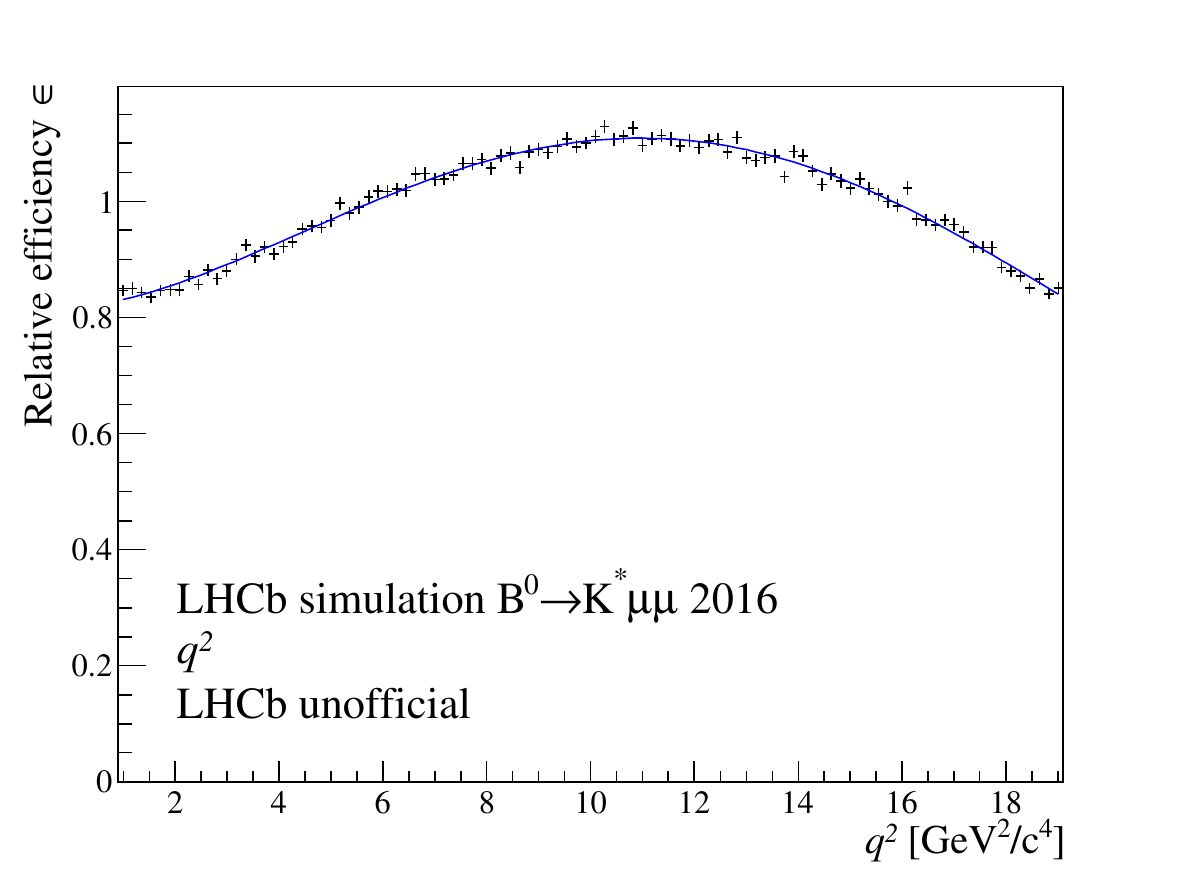}}
\end{minipage}
\vspace*{-2mm}
\caption{\label{fig:acc} One-dimensional acceptance projections for $B^{0}\rightarrow K^{*0} \mu \mu$ decays in $0.4<q^2<19.1~GeV^4/c^2$ determined on \lhcb simulation for the year 2016. The plots show the $cos(\theta_K)$ (left) and $q^2$ (right) projections of the parameterization along with simulated events. }
\end{figure}

The decay \BdToKstmm can be fully described by the three decay angles \thetal, \thetak, $\phiangular$, the invariant mass of the $K\pi$ system, $m_{K\pi}$, and the squared invariant mass of the $\mu^+ \mu^-$ system, \qsq.
This analysis aims to measure the full set of angular observables and the branching fraction in bins of \qsq.
The normalized differential decay rate used to fit the data is given by
\begin{equation}
\renewcommand{\arraystretch}{1.2}
\begin{array}{rc@{\,}c@{\,}l}
    \frac{1}{d(\Gamma+\overline{\Gamma})/dq^2}
    \frac{   \Gamma  }{d\vec{\Omega} \ dq^2 \ dm_{K\pi} } =  &\tfrac{9}{64\pi}&&  \biggl( \sum_{i \in \mathcal{P}}  ({S_i} \pm {A_i}) f_i(\theta_{l},\theta_{K},\phi) {   |\mathcal{BW_{P}}(m_{K\pi})|^{2}  }  \\
    &+&& \sum_{i \in \mathcal{S}} (S_i \pm {A_i}) f_i(\theta_{l},\theta_{K},\phi) {  |\mathcal{L_{S}}(m_{K\pi})|^{2} }  \\
    &+&& \sum_{i \in \mathcal{S/P}}  ({S_i} \pm {A_i}) f_i(\theta_{l},\theta_{K},\phi) {  g_i(\mathcal{L_{S}}(m_{K\pi})\mathcal{BW_{P}^{\star}}(m_{K\pi})) } \biggr).
    
\end{array}
\label{Eq:decayrate}
\end{equation}
The \mkpi dependence of the decay rate is included directly in the angular fit PDF.
The precise description of the \mkpi shape is essential to disentangle spin-1 (P-wave) and spin-0 (S-wave) contributions to the $K\pi$ system. 
The P-wave resonance (\Kstarz) peaks in the \mkpi distribution and is parameterized using a relativistic Breit-Wigner function ($\mathcal{BW_{P}}$).
The P-wave part of the differential decay rate is described by the first term in Eq.~\ref{Eq:decayrate}, while the second term represents the S-wave part. The third term arises from the interference between P- and S-wave.
S-wave contributions from $\Kstarz(1430)$ and $\kappa(800)$ decays exhibit a broad shape in the $m_{K\pi}$ spectrum and are described by the LASS model~\cite{LASS} ($\mathcal{L_{S}}$). 
The function $g_i$ in the third term returns either the $\ensuremath{\mathcal{R}e\xspace}$ or $\ensuremath{\mathcal{I}m\xspace}$ part of its argument depending on the index i.  
The five-dimensional angular fit PDF depends on the three decay angles, \mkpi, and the invariant mass of the \Bd candidate \mkpimumu.
\mkpimumu specifically is used to distinguish signal candidates, which peak around the \Bd mass, from the remaining combinatorial background.
This type of background arises from randomly combined tracks in the detector and is distributed exponentially in \mkpimumu. 
The background shape in \mkpi and the decay angles is modeled using polynomial functions.
A five-dimensional maximum likelihood fit is performed in each \qsq bin separately to extract the CP-symmetric ($S_i$) and CP-asymmetric ($A_i$) angular observables and their uncertainties.
The \qsq spectrum from $0.06\gevgevcccc$ and $19.0\gevgevcccc$ is analyzed in 8 narrow bins of an approximate width of $2\gevgevcccc$. Additionally, the angular observables are extracted in two wide bins from $1.1\gevgevcccc$ to $6.0\gevgevcccc$ and $15.0\gevgevcccc$ to $19.0\gevgevcccc$.
A wide range in \mkpi is chosen to include both the \Kstarz resonance as well as a significant S-wave component.
The spectrum of the decay angles, \qsq and \mkpi in data is distorted by the selection and reconstruction of the decays.
In order to take this acceptance effect into account in the fit to data, it is modeled using five-dimensional Legendre polynomials.
To this end, the polynomial parameters are extracted from fully reconstructed and selected \lhcb simulation of \BdToKstmm decays using the method of moments~\cite{moments}.
Separate acceptance functions are calculated for \Bd and \Bdb decays in order to take detection asymmetries into account.
Example projections of the resulting parameterizations are displayed in Fig.~\ref{fig:acc}. 
The parameterization is either evaluated at a fixed point in \qsq and used as a multiplicative factor in the probability density function or its inverse value is used as an acceptance correction weight.
The first method is used in the narrow \qsq bins, where the variation of the acceptance across the \qsq bin is small and the evaluation of the acceptance at a single point is valid.
In bins where this is not the case, a weighted fit is performed.

\section{Improvements to analysis strategy}

Compared to the previous binned measurement~\cite{LHCb-PAPER-2020-002}, the analysis strategy has been improved in several ways.
The description of the \mkpi dependence is now directly included in the angular PDF as described in Eq.~\ref{Eq:decayrate}.
Previously, the \mkpi distribution has only been fitted simultaneously to the angular fit.
Along with the wider \mkpi selection window compared to the previous analysis, this significantly reduces the uncertainty on the observables describing the P/S-wave interference, which will be measured for the first time.
The selection has been improved to increase the signal efficiency and background rejection by re-optimizing the boosted decision tree against combinatorial background and the particle identification selection against misidentified physical backgrounds.
As displayed in Eq.~\ref{Eq:decayrate}, the CP-asymmetries are extracted simultaneously with the CP-symmetries.
This is an improvement compared to previous iterations, where the asymmetries have either not been extracted or have only been determined separately from the CP-symmetries.
Measuring the asymmetries simultaneously in addition preserves the correlations between the symmetries and asymmetries.
To constrain the total CP-asymmetry \ACP an extended term is added to the likelihood.
The information from the extended term is also used to measure the branching fraction, where the relative signal efficiency can be determined model-independently using information from the angular observables and the acceptance. 

\begin{figure}
    \centering
    \includegraphics[width=0.55\textwidth]{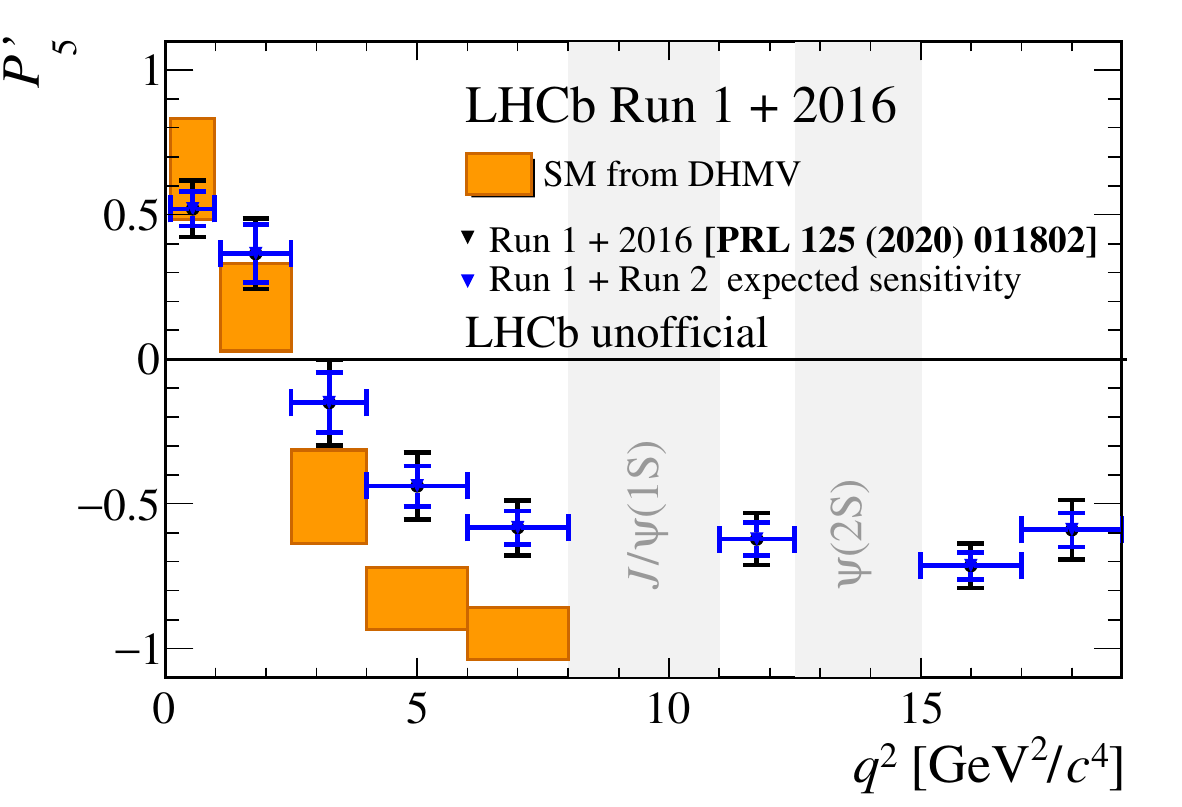}
    \vspace*{-2mm}
    \caption{ Measurement of the less form-factor dependent observable $P_5^\prime$ (left) by \lhcb using Run~1 and 2016 data in black. The projected total uncertainty using the full Run~1 and Run~2 \lhcb data sample assuming identical central values is shown in blue.
    \label{fig:P5p_improvement}}
\end{figure}

Overall the sensitivity to the angular observables is expected to increase significantly compared to the previous iteration of the analysis.
The major part of the improvement originates from the inclusion of 2017/18 \lhcb data samples, which effectively double the number of signal candidates.
Furthermore, the statistical uncertainty is reduced by the re-optimization of the selection and adding the \mkpi dependence directly to the angular PDF.
Figure~\ref{fig:P5p_improvement} shows the result of the previous binned angular analysis of \BdToKstmm decays by \lhcb using Run~1 and 2016 data for the optimized observable $P_5^{\prime}$ (black).
Overlayed in blue the same result with the projected total uncertainty using the full Run~1 and Run~2 data sample is shown.

\section{Conclusion}
In conclusion, the update of the binned angular analysis using the full Run~1 and Run~2 \lhcb data sample is expected to significantly improve the precision of the measurement of the binned angular observables. The observables describing the angular structure of the S-wave and interference between S- and P-wave will be published for the first time.
Furthermore, this analysis will publish the most complete set of the angular asymmetries of the decay \BdToKstmm to date along with a model-independent measurement of the branching fraction.

\section*{References}

\begin{thebibliography}{99}



\bibitem{LHCb-PAPER-2020-002}R. Aaij et al. [LHCb], \Journal{\PRL}{125}{011802}{2020}, \href{https://arxiv.org/abs/2003.04831}{[arXiv:2003.04831]}.

\bibitem{kstplus_angular}R. Aaij et al. [LHCb],\Journal{\PRL}{126}{161802}{2021}, \href{https://arxiv.org/abs/2012.13241}{[arXiv:2012.13241]}.
 
\bibitem{phi_angular} R. Aaij et al. [LHCb], JHEP \textbf{11} (2021) 043, \href{https://arxiv.org/abs/2107.13428}{[arXiv:2107.13428]}.



\bibitem{isospin}R. Aaij et al. [LHCb], JHEP \textbf{06} (2014) 133, \href{https://arxiv.org/abs/1403.8044}{[arXiv:1403.8044]}.

\bibitem{swave-paper}R. Aaij et al. [LHCb], JHEP \textbf{04} (2017) 142, \href{https://arxiv.org/abs/1606.04731}{[arXiv:1606.04731]}.

\bibitem{bsphimumu} R. Aaij et al. [LHCb], \Journal{\PRL}{127}{151801}{2021}, \href{https://arxiv.org/abs/2105.14007}{[arXiv:2105.14007]}.

\bibitem{lambda} R. Aaij et al. [LHCb], JHEP \textbf{09} (2018) 145, \href{https://arxiv.org/abs/1503.07138}{[arXiv:1503.07138]}.


\bibitem{optimized_observables}S. Descotes-Genon et al.  JHEP \textbf{01} (2013) 048, \href{https://arxiv.org/abs/1207.2753}{[arXiv:1207.2753]}

\bibitem{amplitudeanalysis}R. Aaij et al. [LHCb],\Journal{\PRL}{132}{131801}{2024}, \href{https://arxiv.org/pdf/2312.09115}{[arXiv:2312.09115]}.

\bibitem{amplitudeanalysis2}R. Aaij et al. [LHCb],\Journal{\PRD}{109}{052009}{2024}, \href{https://arxiv.org/pdf/2312.09102}{[arXiv:2312.09102]}.

\bibitem{dispersion_relation}R. Aaij et al. [LHCb], LHCb-PAPER-2024-011, in preparation

\bibitem{Descotes-Genon} Descotes-Genon et al., JHEP \textbf{12} (2014) 125, \href{https://www.arxiv.org/abs/1407.8526}{[arXiv:1407.8526]}.


\bibitem{Khodjamirian:2010vf} A. Khodjamirian et al., JHEP \textbf{09} (2010) 089, \href{https://arxiv.org/abs/1006.4945}{[arXiv:1006.4945]}.

\bibitem{LASS}D. Aston et al., Nucl. Phys. B \textbf{296} (1988) 493.

\bibitem{moments} W. Eadie et al., Statistical methods in experimental physics, Journal of the American Statistical Association (2013).




\end{thebibliography}
\end{document}